\documentclass[twocolumn]{aastex701}

\def\simgr{\,\hbox{\hbox{$ > $}\kern -0.8em \lower 1.0ex\hbox{$\sim$}}\,}
\def\simle{\,\hbox{\hbox{$ < $}\kern -0.8em \lower 1.0ex\hbox{$\sim$}}\,}

\newcommand{\xmm}{XMM-Newton}

\newcommand{\mspfive}{4FGL J1646.5$-$4406}
\newcommand{\mspfour}{PSR J1311$-$3430}
\newcommand{\mspthree}{4FGL~J1838.2+3223} 
\newcommand{\msptwo}{PSR~J1048$+$2339}
\newcommand{\mspone}{PSR~J0838$-$2827}
\newcommand{\fgl}{3FGL~J0838.8$-$2829}
\newcommand{\obj}{1FGL~J0523.5$-$2529}

\shortauthors{Halpern \& Bogdanov}
\shorttitle{\xmm\ Observation of \obj}

\begin{document}
\title{\xmm\ Observation and Optical Monitoring of the Candidate Redback Millisecond Pulsar \obj}

\author[0000-0003-4814-2377]{Jules P. Halpern}
\affiliation{Department of Astronomy, Columbia University, 550 West 120th Street, New York, NY 10027-6601, USA}
\email {jph1@columbia.edu}
\author[0000-0002-9870-2742]{Slavko Bogdanov}
\affiliation{Columbia Astrophysics Laboratory, Columbia University, 550 West 120th Street, New York, NY 10027-6601, USA}
\email {slavko@astro.columbia.edu}

\begin{abstract}

  \obj\ is a Fermi selected redback millisecond pulsar candidate that exhibited luminous optical and X-ray flares in 2020--2021.  We obtained a simultaneous X-ray and $U$-band observation with \xmm\ in 2025, the first to cover the 16.5~hr orbit of \obj.  The X-ray luminosity was in an intermediate state with a power-law photon spectral index of $\Gamma=1.53\pm0.02$.  Apparent fluctuations were superposed on a broad, single-peaked modulation, the latter characteristic of intrabinary shock models in which the shock front is wrapped around the pulsar.  The $U$-band light curve was dominated by ellipsoidal modulation of the nearly Roche lobe filling companion star, similar to that seen in ground-based optical photometry.  We also used this effect in 10 years of ATLAS monitoring to improve the precision of the orbital period to 0.6881366(19) days.  Considering that searches for radio pulsations from \obj\ at all orbital phases have been unsuccessful, it may be that the shocked wind usually surrounds the pulsar.
  
\noindent  
{\it Unified Astronomy Thesaurus concepts:} Millisecond pulsars (1062); Binary pulsars (153); Radio pulsars (1353); Rotation-powered pulsars (1408); Pulsars (1306); Neutron stars (1108)
\end{abstract}

\section{Introduction\label{sec:intro}}

The majority of millisecond pulsars (MSPs) have either a white dwarf or a low-mass,
evolved star as a companion.  The canonical explanation for the origin of MSPs is
``recycling'' of the neutron star by accretion during a low-mass
X-ray binary (LMXB) phase \citep{alp82,rad82}, after which radio pulsations turn
on because accretion has stopped.  A highlight of the Fermi mission is its
discovery of many $\gamma$-ray-emitting black widows (BWs) and redbacks,
previously rare radio MSPs
in which the relativistic pulsar wind heats and ablates the photosphere of
a companion star.  BWs have orbital periods $\le1$~day and degenerate
companions of $<0.05\,M_{\odot}$, while redbacks have non-degenerate
$0.1-1\,M_{\odot}$ companions that are generally hotter and less dense
than main-sequence stars of the same mass. \citet{kol25} cataloged
50 BWs and 30 redbacks in the Galactic field (not including those
in globular clusters).  In addition, they list five BW candidates and
23 redback candidates from which pulsations have not yet been found.

Redback companions are close to filling their Roche lobes \citep{str19}.
Their winds can obscure the radio pulsar signal for a large fraction
of the orbit (e.g., \citealt{den21,per23,tho24}), and sometimes produce
optical emission lines.  Nonthermal emission from redbacks is
attributed to shocks from the collision between the pulsar and
stellar winds and is dominant in X-rays \citep{aln21,bog05,bog11,bog14}.
X-ray flux is usually modulated around the orbit, which is
interpreted in terms of the geometry of the intrabinary shock,
with synchrotron emission beamed by particles moving tangentially
to the shock surface \citep{rom16,wad17,wad18,kan19,van20,cor22,cor24,cor25}.
  
The photosphere of the companion can be heated by the high-energy
photons produced by the shock or directly by pulsar wind particles
\citep{san17}.  The optical modulation observed around
the orbit is sometimes dominated by this heating effect. Alternatively,
when heating is weak, intrinsic ellipsoidal modulation of light 
from the tidally distorted, nearly Roche-lobe-filling star is the main
contributor \citep{li14,bel16,van16,sha17}, with large starspots also
possibly playing a role.  All of these effects can vary. Photospheric
heating patterns change over weeks and months.  In several redbacks
minutes-long flares increase the luminosity by a factor of 10--100 in X-rays,
and up to $\sim1$ magnitude in optical, e.g., in \msptwo\ \citep{cho18},
\mspone\  \citep[=\fgl:][]{hal17,rea17}, and \mspthree\ \citep{zyu24}.

The subject of this paper is \obj, the first candidate redback to be
identified \citep{str14}, in this case from an X-ray counterpart in
Swift and an optical light curve consistent with pure elliptical modulation
with a period of 15.6~hr in the Catalina Real-Time Transient Survey data
(CRTS; \citealt{dra09}).
Radio pulsations have not been detected from this source despite extensive
searches \citep{tho24,joh25}.  Spectroscopy of the early K star measured its
orbital velocity and rotational broadening \citep{str14}, indicating a mass 
ratio of $0.61\pm0.06$ and a companion star mass in the 
range $0.8-1.3\,M_{\odot}$, the largest among redbacks. These parameters
require an orbital inclination in the range $57^{\circ}<i<80^{\circ}$.
We assume a distance of 2.24~kpc from the Gaia measured parallax of
$0.446\pm0.042$~mas in its extended third data release (EDR3; \citealt{bro21}).

In early 2020, we discovered dramatic optical flaring from \obj\ and carried out a Swift campaign \citep[][hereafter Paper~1]{hal22} that detected simultaneous X-ray and optical/UV flares.  Flares were also associated with the emergence of optical emission lines.  The flares had rise times of a few minutes, sometimes recurred for hours or days, and continued intermittently through 2021.  A wide range of X-ray luminosity was observed during the campaign, spanning a factor of $\approx100$ between a barely detected minimum and the peak of the brightest flare. 

Here we present continued ground-based optical monitoring through early 2026, together with results from an \xmm\ pointing, the first X-ray observation sensitive to the non-flaring state and to cover a full 16.5~hr binary orbit.   Section~\ref{sec:opt} presents the optical time-series data from the MDM Observatory and monitoring data from the Zwicky Transient Facility (ZTF; \citealt{bel19}) and the Asteroid Terrestrial-impact Last Alert System (ATLAS; \citealt{ton18}). The \xmm\ observation is described in Section~\ref{sec:xmm}. The results are discussed and interpreted in Section~\ref{sec:disc}. The main conclusions are summarized in Section~\ref{sec:conc}.

\section{Optical Observations\label{sec:opt}}

\subsection{MDM Observatory 1.3 m and ZTF}

\begin{deluxetable}{lcrcc}
\label{tab:optlog}
\tablecolumns{4} 
\tablewidth{0pt} 
\tablecaption{Log of MDM $r$-band Time-series Photometry}
\tablehead{
\colhead{Date} & \colhead{Exp.} & $N_{\rm exp}$ &
\colhead{Time} & \colhead{Phase} \\
\colhead{(UT)} & \colhead{(s)} & & \colhead{(UTC)} &
\colhead{($\phi$)}
}
\startdata
2023 Jan 23  & $60$ &  68 & 01:40--02:54 & 0.991--0.066 \\
2023 Jan 25  & $60$ & 185 & 01:40--04:53 & 0.897--0.093 \\
2023 Jan 28  & $60$ &  85 & 01:40--03:09 & 0.257--0.347 \\
\hline
2023 Nov  7  & $60$ & 366 & 06:32--12:58 & 0.809--0.198 \\
2023 Nov  8  & $60$ & 366 & 06:31--12:57 & 0.261--0.650 \\
2023 Nov  9  & $60$ &  99 & 06:24--08:07 & 0.707--0.811 \\
2023 Nov 10  & $60$ & 174 & 06:20--09:22 & 0.155--0.339 \\
2023 Nov 11  & $60$ & 171 & 06:14--09:13 & 0.603--0.783 \\
\hline
2024 Nov  5  & $60$ & 341 & 06:36--12:34 & 0.777--0.138 \\
2024 Nov  6  & $60$ & 265 & 06:41--11:19 & 0.235--0.516 \\
2024 Nov  7  & $60$ & 347 & 06:53--12:58 & 0.700--0.069 \\
2024 Nov  8  & $60$ & 357 & 06:22--12:38 & 0.123--0.502 \\
\hline
2025 Jan 26  & $60$ &  85 & 02:49--04:17 & 0.709--0.798 \\
2025 Jan 27  & $60$ & 148 & 01:39--04:14 & 0.092--0.248 \\
2025 Jan 28  & $60$ & 117 & 01:36--03:38 & 0.542--0.665 \\
\hline
2025 Sep 29  & $60$ & 197 & 09:10--12:36 & 0.579--0.787 \\
\hline
2025 Dec 20  & $60$ & 367 & 03:44--10:09 & 0.416--0.805 \\
2025 Dec 21  & $60$ & 365 & 03:46--10:09 & 0.871--0.258 \\
\hline
2026 Jan 20  & $60$ & 369 & 01:41--08:10 & 0.339--0.733 \\
2026 Jan 21  & $60$ & 344 & 01:58--08:00 & 0.811--0.175 \\
\enddata
\end{deluxetable}

\begin{figure*}
  \centerline{
\includegraphics[trim={0.8cm 4.6cm 0.8cm 5.6cm},clip,angle=0.,width=1.0\linewidth]{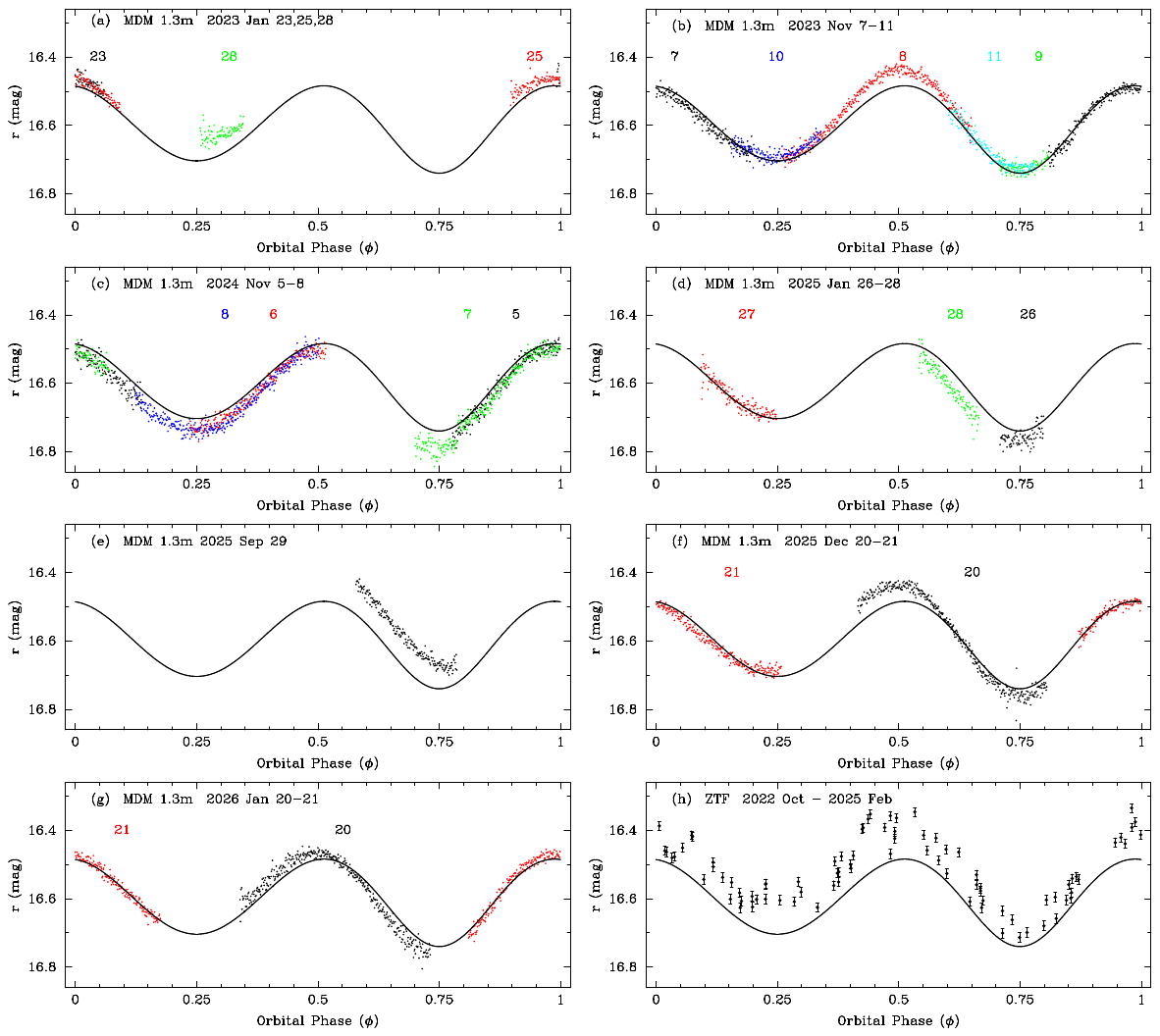}
}
\caption{Panels a--g: $r$-band time-series photometry of \obj\ during seven runs on the MDM 1.3~m from 2023 to 2026, folded according to the orbital ephemeris described in Section~\ref{sec:ephem}.  A log of these observations is given in Table~\ref{tab:optlog}.  The data points are color coded and labeled by the day of the month.   A fixed model of ellipsoidal modulation (not a fit to the data) is superposed as an aid to visualizing deviations and variability. Panel h is the ZTF $r$-band data from an overlapping period, which is consistent with the MDM data except for an apparent $\approx 0.07$ mag offset in calibration.
}
\label{fig:time-series}
\end{figure*}

\begin{figure*}
  \centerline{
\includegraphics[trim={0.8cm 4.6cm 0.8cm 5.6cm},clip,angle=0.,width=1.0\linewidth]{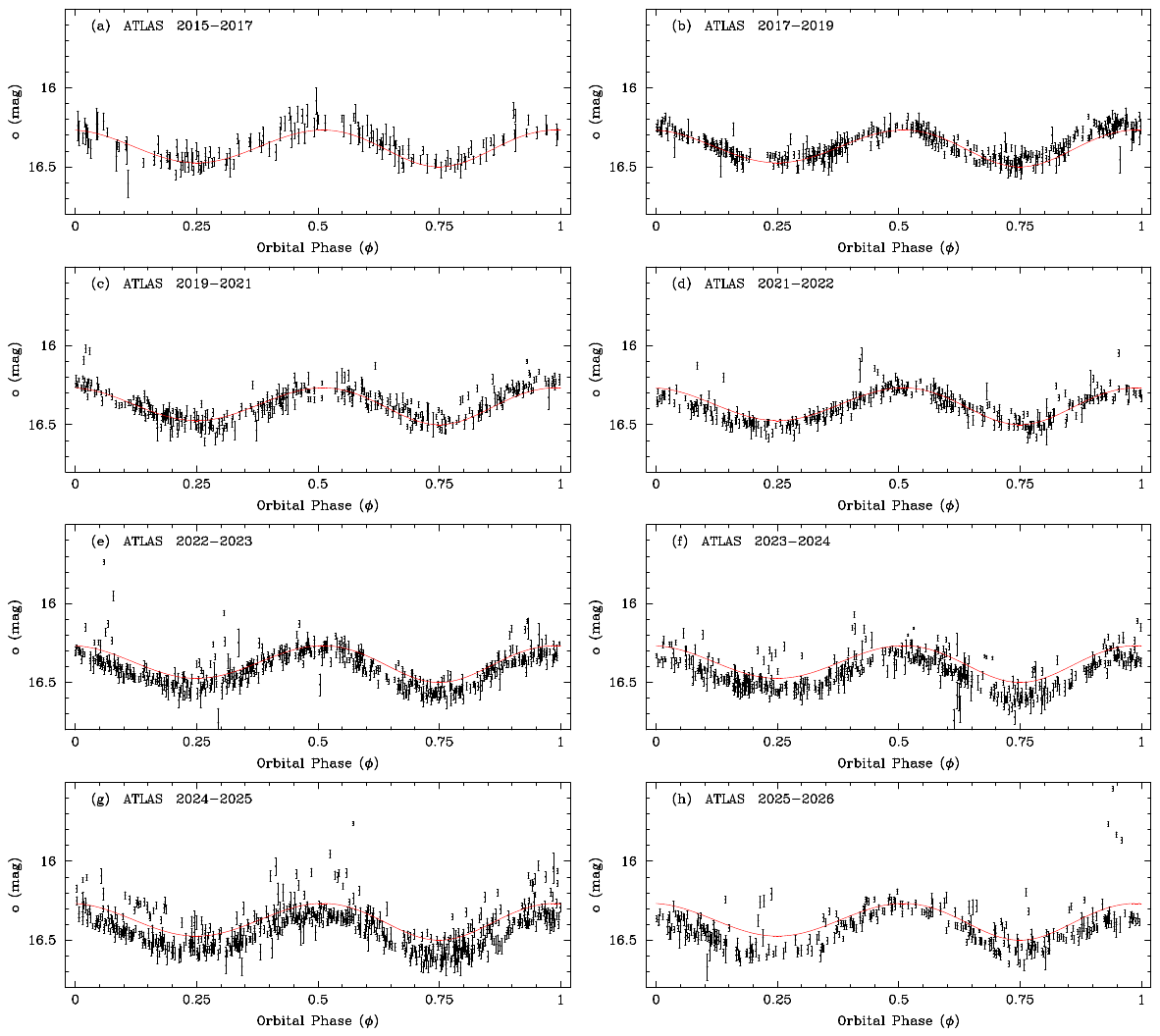}
}
\caption{ATLAS forced photometry in the $o$ filter folded according to the orbital ephemeris described in Section~\ref{sec:ephem}.   Each panel contains either one or two observing seasons.  A fixed model of ellipsoidal modulation (not a fit to the data) is superposed as an aid to visualizing deviations and long-term variability.  
}
\label{fig:atlas}
\end{figure*}

We used the MDM Observatory 1.3~m McGraw-Hill telescope on Kitt Peak for time-series photometry of \obj\ in the $r$ band for 20 nights over seven observing runs in 2023--2026, as listed in Table~\ref{tab:optlog}.  The methods are identical to those used in Paper~1.  A thinned, backside-illuminated SITe CCD (Templeton) was windowed to achieve an efficient read/prep cycle time of 3~s in between 60~s exposures.  The times were converted to TDB using the online tool\footnote{\url{https://astroutils.astronomy.osu.edu/time/utc2bjd.html}} of \citet{eas10} and the Gaia-CRF3 position R.A.=$05^{\rm h}23^{\rm m}16.\!^{\rm s}931$, decl.=$-25^{\circ}27^{\prime}37.\!^{\prime\prime}13$ (\citealt{bro21}), which is referenced for proper motion to epoch 2016.0.

  Differential photometry was performed with respect to the PanSTARRS magnitude of a nearby comparison star, the long-term stability of which was verified in ATLAS and ZTF data. Figure~\ref{fig:time-series} shows the resulting photometry folded using the orbital ephemeris described in Section~\ref{sec:ephem}.  Due to the 16.5~hr orbital period and southerly declination of \obj, it is possible to cover at most 0.4 cycles per night from Kitt Peak; a five-night span, which was rarely available, is needed to sample all phases.

As in Paper~1, an analytic model of ellipsoidal modulation \citep{mor93, gom21} is superposed on the light curves.  The orbital inclination $i = 67^{\circ}$ and the mass ratio $q=0.61$ come from the spectroscopic study by Strader et al. (2014), while Roche-lobe filling factor $f = 0.99$, linear limb-darkening coefficient 0.65, and gravity-darkening coefficient 0.45 are assumed.  The two coefficients are suitable for an atmosphere of $T_{\rm eff}=5,000$~K and log~$g$ = 4.0 \citep{cla11,tor21}. The fixed model curve is not a fit to the data, but a fiducial to illustrate the variety of significant observed deviations from the model (see Paper 1 for details), which might be attributable to asymmetric and variable photospheric heating of the companion star's photosphere by the pulsar and intrabinary wind shock. We also display 85 $r$-band data points in 2022--2025 from the ZTF 24th data release, which largely overlap in time with the MDM observations and match the ellipsoidal model to first order except for a possible $\approx0.07$~mag offset in calibration.

\subsection{ATLAS\label{sec:atlas}}

We downloaded ATLAS forced photometry of \obj, which is the densest long-term optical record of this source.  After applying conservative quality filters\footnote{\url{https://fallingstar-data.com/forcedphot/faq/}} and removing points with uncertainties $>0.1$ mag, there are 3522 observations in the ``orange'' $o$ filter (560--820 nm) and 946 points in the ``cyan'' $c$ filter (420--650 nm) spanning 2015 October 8 to 2026 January 17.  The $o$ points in Figure~\ref{fig:atlas} are grouped into one or two observing seasons per panel.  The same ellipsoidal model as above is used as a fiducial marker of long-term variability.  Gradual fading by $\approx0.15$ mag over 10 years is seen in both the $o$ and $c$ filters.  The same trend is present in the ZTF $r$ and $g$ filters (from 2018 to 2025, not shown here), with a slight increase in $g-r$.

There are a number of outlying points that may be indicative of flares such as those documented in Paper~1 from high-cadence MDM data.  Because ATLAS takes four 30~s exposures of each field within a 1 hr period, flares may be confirmed by more than one detection in the sequence of four.  We have not examined all of the high points in detail, but a notable example of a probable flare occurs in Figure~\ref{fig:atlas}h, where all four high points around phase 0.95 were taken over a span of 27~minutes beginning on 2025 November 29 04:16:58 UT.

\subsection{Updated Orbital Ephemeris\label{sec:ephem}}

The original spectroscopic ephemeris developed in 2013--2014 \citep{str14} is still the only one published: $P=0.688134\pm0.000028$~day, $T_{0.75}=2456577.64636\pm0.0037$~(BJD).  Here $T_{0.75}$ is the epoch of superior conjunction of the companion star, which occurs at phase $\phi=0.75$ in the radio pulsar convention, where $\phi=0$ is the ascending node of the pulsar.  The times of extrema of the optical light curves from 2020--2021 in Paper~1 were adequately matched by the simple ellipsoidal model using this ephemeris. Nevertheless, considering the significant deviations of the light curve from the simple model, there is a systematic ambiguity in the phase of the photometric extrema relative to the spectroscopic epoch that is greater than the $\approx 5$ minute precision of the latter.  Therefore, we choose the spectroscopic (kinematic) epoch as the fiducial point to define the phases of the X-ray and optical light curves.  However, the uncertainty of the original spectroscopic orbital period extrapolates to a large expected error of $\approx 0.18$ cycles at the epoch of the \xmm\ observation.  

Fortunately, the 10~yr span of the ATLAS data set affords a significant improvement in the precision of the orbital period.  For this purpose, we applied a Lomb-Scargle periodogram to the data in Figure~\ref{fig:atlas}, excluding points with $o\le16.1$ or $o>16.7$.  Noting that nearly all power is in the first harmonic due to the double-peaked wave form, we use the harmonic to get $P=0.6881366(19)$~days for the orbital period, an order-of-magnitude refinement that is fortuitously close to the original value.  This is the average period over the 10-yr span. Although it is known from radio and $\gamma$-ray pulse timing that the orbital periods of redbacks change stochastically on timescales of years, this only appears as drifts in the time of ascending node by tens of seconds (e.g., \citealt{den16,tho24}), an effect that is too small to be detected in the photometry. Adopting here a hybrid ephemeris consisting of the spectroscopic epoch and the ATLAS period, the propagated uncertainty in the phasing of the X-ray and optical light curves relative to the kinematic orbit is acceptably $\approx0.0126$ days (0.018 cycles). 

\section{\xmm\ Observation\label{sec:xmm}}

We observed \obj\ with \xmm\ for $61$~ks on 2025 February 25.  The EPIC pn \citep{2001A&A...365L..18S}, MOS1, and MOS2 \citep{2001A&A...365L..27T} cameras were used in the large window mode with the thin optical blocking filter.  All three data sets were processed with the Science Analysis Software (SAS) package version 22.1.0.  The standard flag and pattern event screening criteria were applied.  The \xmm\ Optical Monitor (OM; \citealt{2001A&A...365L..36M}) was used with the $U$ filter and a fast-mode window with a tracking frame duration of 20~s.  A photometric light curve was extracted using the {\tt omfchain} processing pipeline with the default parameter settings.

\begin{figure*}
  \centerline{
\includegraphics[angle=0.,width=0.8\linewidth]{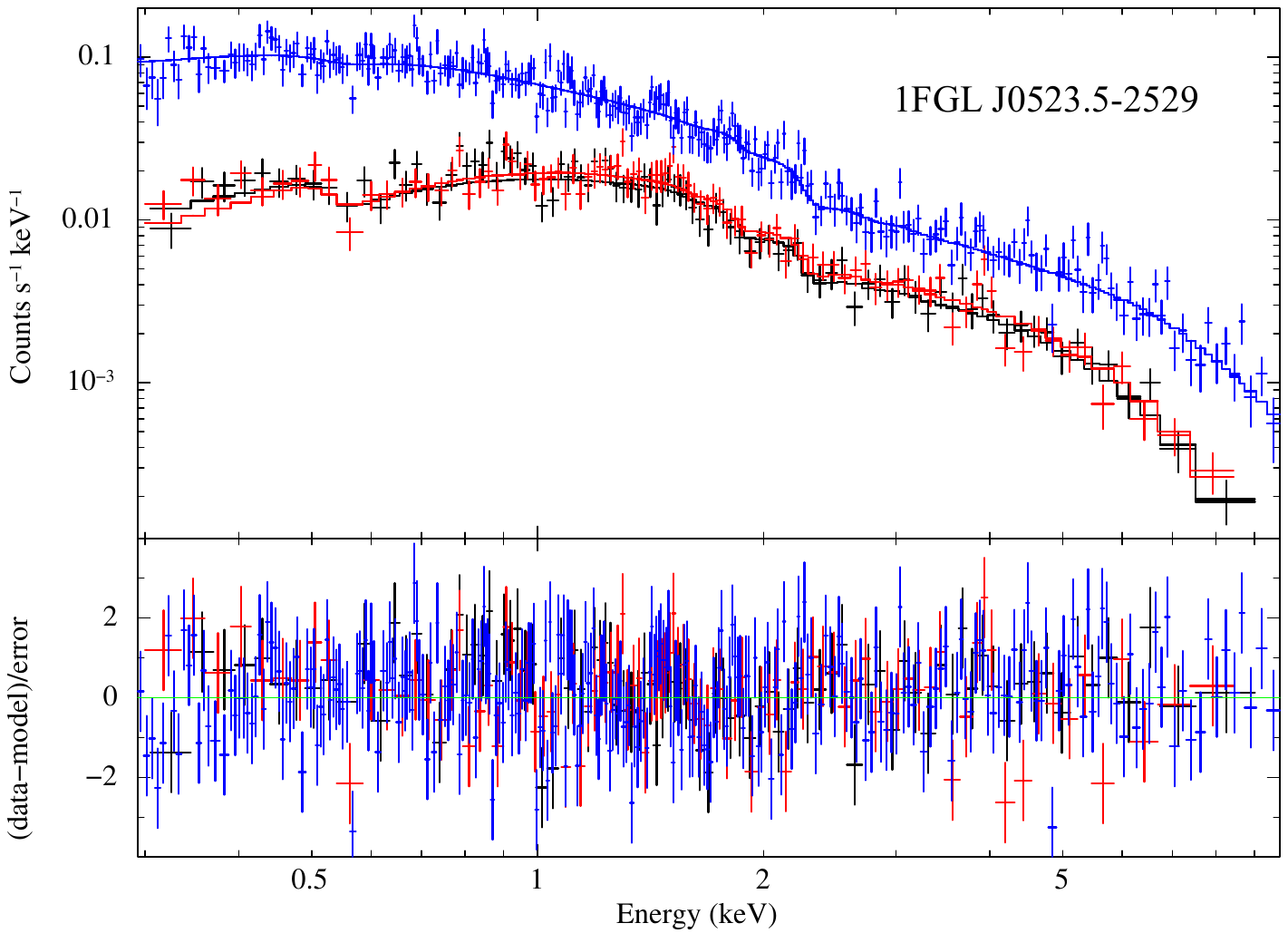}
}
\caption{Total \xmm\ spectrum and power-law fit, with pn data in blue and MOS in black and red. The bottom panel shows the residuals from the best fit. 
}
\label{fig:spec}
\end{figure*}

\begin{figure*}
  \centerline{
\includegraphics[trim={0.0cm 6.0cm 0.0cm 6.6cm},clip,angle=0.,width=0.8\linewidth]{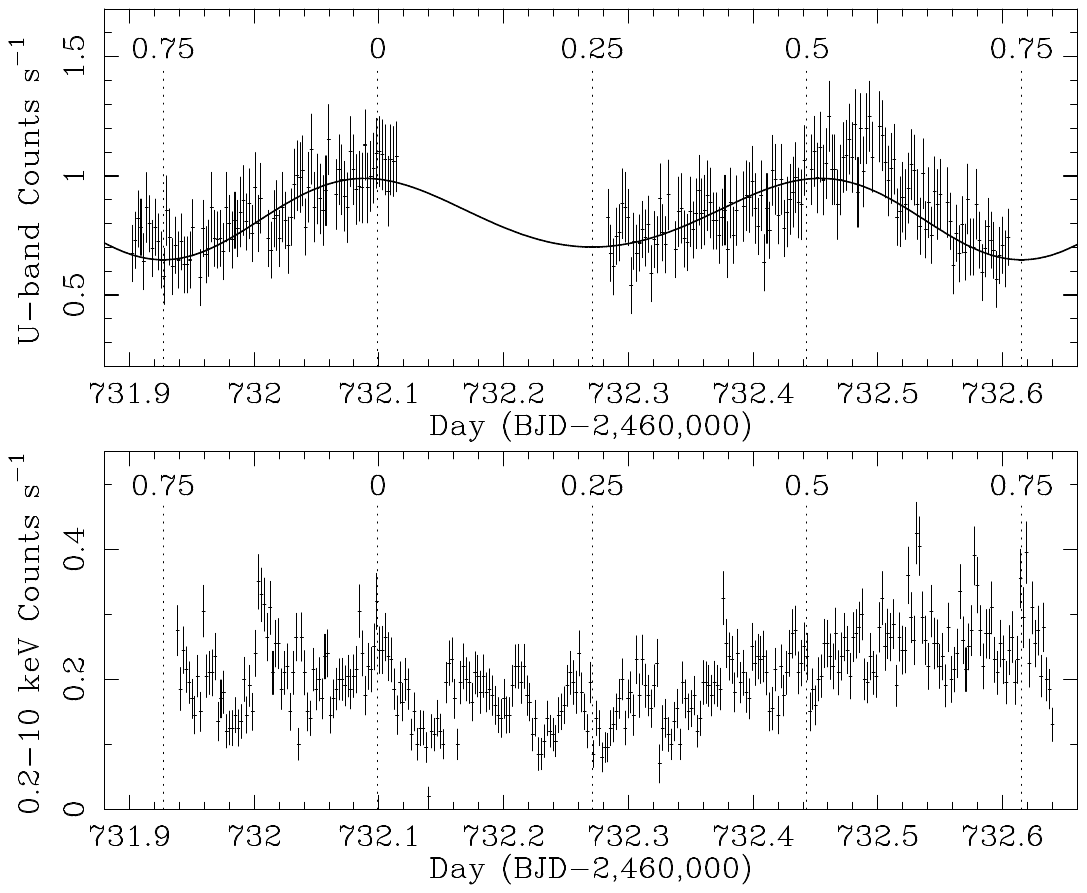}
}
  \caption{\xmm\ light curves in 200~s bins, with orbital phases marked. Top: The gap in the OM data is due to a telemetry drop that caused a telecommand failure of three of the 13 OM observations.  Otherwise, the light curve is broadly consistent with the ellipsoidal model except perhaps for some increase around day 732.5.  Bottom: Background subtracted and summed X-ray count rates from the pn and MOS detectors showing a broad minimum around phase 0.25 and some fluctuations.
}
\label{fig:xmm}
\end{figure*}

\begin{figure}
\centerline{\hspace{3.2in}
\includegraphics[angle=0.,width=2.2\linewidth]{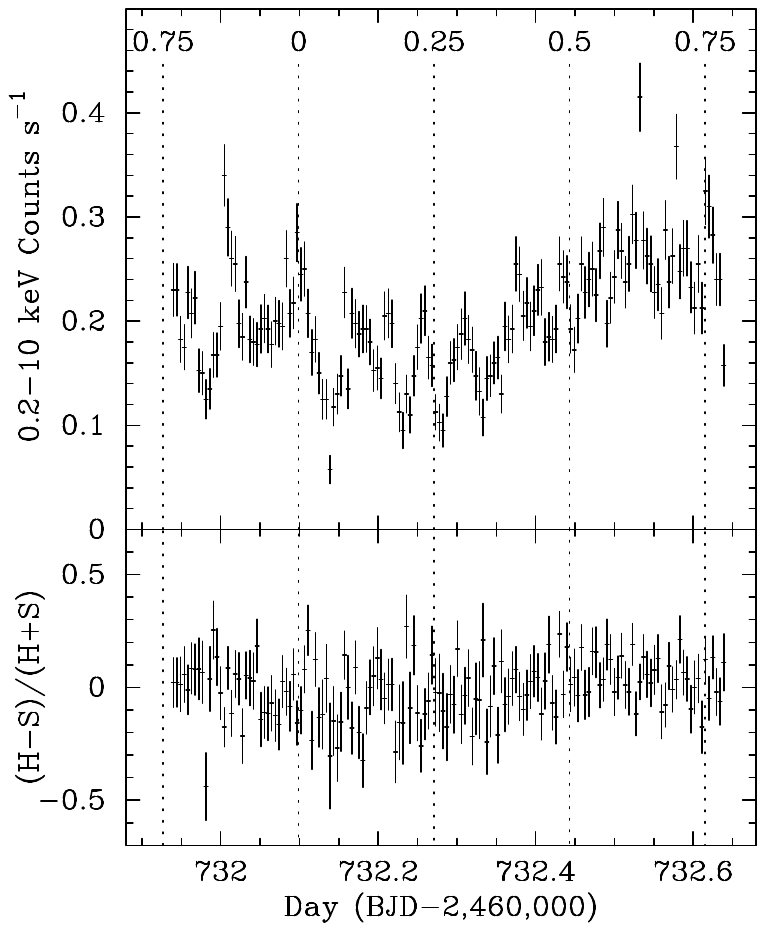}}
\vspace{-1.in}
\caption{Top: The X-ray light curve of Figure~\ref{fig:xmm} in 400~s bins, with orbital phases marked.  Bottom: Hardness ratio of the 0.2--1.2 keV and 1.2--10 keV bands with the same binning.
}
\label{fig:binned}
\end{figure}

\begin{deluxetable*}{lccc}
\tabletypesize{\normalsize}
\tablewidth{0pt}
\tablecolumns{4}
\tablecaption{\xmm\ Spectral Fits \label{tab:xspec}}
\tablehead{\colhead{Parameter} & \colhead{} & \colhead{Orbital phase interval} & \colhead{} }
\startdata
    & $0.0-1.0$ & $0.0-0.5$ & $0.5-1.0$ \\
\hline
$N_{\rm H}$ ($10^{20}$ cm$^{-2}$)  & $1.4\pm0.5$  & $1.3\pm0.8$ & $1.8\pm0.6$  \\
$\Gamma$  & $1.53\pm0.02$ & $1.54\pm0.03$ & $1.53\pm0.03$ \\
PL norm. ($10^{-5}$ keV$^{-1}$ cm$^{-2}$ s$^{-1}$) & $6.78\pm0.13$  & $5.38\pm0.16$ & $8.16\pm0.20$ \\
$F_X$\tablenotemark{a} ($10^{-13}$\,erg\,cm$^{-2}$\,s$^{-1}$) & $5.48\pm0.07$ & $4.33\pm0.09$ & $6.60\pm0.11$\\
$\chi_{\nu}^2$/DoF  & $1.10/478$ & $1.17/211$  &  $1.04/297$ \\
\enddata
\tablenotetext{a}{Unabsorbed flux in the 0.3--10 keV range.}
\end{deluxetable*}

\subsection{X-ray Spectrum\label{sec:spec}}

For spectral analysis, source events were extracted from a circular region of radius 30$''$ centered on the Gaia-CRF3 position. Background estimates were obtained from a source free region of radius 60$''$ in the vicinity of the source. The extracted pn and MOS1/2 source spectra were grouped such that each energy bin contains at least 25 counts.
We account for absorption due to the interstellar medium using the \texttt{tbabs} model in XSPEC, which considers the \texttt{VERN} cross-sections \citep{verner96} and \texttt{WILM} abundances \citep{Wilms00}. We consider an absorbed power-law spectrum, which is expected from intra-binary shock emission.

The results of the spectroscopic analysis are summarized in Table~\ref{tab:xspec} and are displayed in Figure~\ref{fig:spec}.  The best fit to the time-averaged spectrum gives an $N_{\rm H}=(1.4\pm0.5)\times10^{20}$ cm$^{-2}$, a photon index $\Gamma=1.53\pm0.02$, and an unabsorbed flux $F_X=(5.48\pm0.07)\times10^{-13}$ erg cm$^{-2}$ s$^{-1}$ in the 0.3--10 keV band, with $\chi_{\nu}^2=1.10$ for 478 degrees of freedom. All errors quoted are at the significance level  $1\sigma$. For a distance of 2.24~kpc, the flux translates into a luminosity $L_X=3.3\times10^{32}$ erg s$^{-1}$. The inferred hydrogen column density is consistent with the value of $1.5\times10^{20}$ cm$^{-2}$ measured through the Galaxy towards \obj\ \citep{hi16}.

To examine any X-ray spectral variations over the binary orbit, we divide the data into two orbital phase segments $0.0-0.5$ and $0.5-1$ and fit a power-law spectrum to each separately. There is no appreciable change in the photon index between the two portions of the orbit (see Table~\ref{tab:xspec}).

\subsection{X-ray and OM Light Curves\label{sec:curve}}

Background-subtracted light curves in the 0.2--10 keV range were created from each EPIC detector by extracting the source counts in a circle of radius $20^{\prime\prime}$ and getting background counts from a circle of the same radius in a neighboring source-free region.  The three light curves were merged in 200~s bins (Figure~\ref{fig:xmm}).  The subtracted background counts are on average 3.8\% of the source counts.  We verified that they do not affect the character of the light curve, which displays some fluctuations superposed on a broad minimum centered around $\phi=0.25$ determined from the ephemeris described in Section~\ref{sec:ephem}.  The X-ray light curve binned by 400~s is shown in Figure~\ref{fig:binned}, together with the corresponding hardness ratio, where $S$ and $H$ are the count rates in the 0.2--1.2 keV and 1.2--10 keV range, respectively.  The hardness ratio is constant across the orbit, consistent with the lack of spectral variability.

The OM light curve in Figure~\ref{fig:xmm} was binned identically to the X-rays.  The gap in the data is due to a telemetry drop that caused a telecommand failure of three of the 13 OM observations.  Otherwise, it is broadly consistent with the superposed ellipsoidal model except perhaps for an excess around day 732.5. The parameters of the ellipsoid model are the same as those used for ground-based photometry except that the limb-darkening and gravity-darkening coefficients are 0.95 and 1.1, respectively, appropriate for the $U$ band \citep{cla11}.  The absence of any UV excess corresponding to the X-ray fluctuations is not unexpected, as discussed in Section~\ref{sec:optres}.

\section{Discussion}
\label{sec:disc}

\subsection{Optical and X-ray Results}
\label{sec:optres}

Bright flares such as those observed in the 2020--2021 MDM $R$-band observations of Paper~1 are not seen in the $r$-band light curves from 2023 to early 2026 presented here in Figure~\ref{fig:time-series}. Nevertheless, there is evidence of occasional flares in the more frequent but brief visits by ATLAS (Figure~\ref{fig:atlas}) during the latter period.  Otherwise, the optical light curve of \obj\ is dominated by ellipsoidal modulation. This also applies to the OM $U$ filter; only around phase 0.57 in Figure~\ref{fig:xmm} is there an apparent excess over the ellipsoidal model.   The zero point of the OM $U$ filter in the AB magnitude system is 19.189 (corresponding to 1 count~s$^{-1}$)\footnote{https://xmmweb.esac.esa.int/docs/documents/CAL-TN-0019.pdf}. Therefore, the observed count rates ranging from 0.65 count~s$^{-1}$ to 1.1 count~s$^{-1}$ convert to $U$=19.66--19.09, which is in excellent agreement with the folded Swift UVOT light curve in 2020 assembled from many brief observations (Paper~1). Only during the brightest flare on 2020 January 20 did the Swift $U$-band magnitude reach 18.02, which would correspond to 2.9 count~s$^{-1}$ in the OM light curve, off the scale in Figure~\ref{fig:xmm}. 

The \xmm\ observation is the first to cover the entire orbit in X-rays with sensitivity sufficient to characterize the light curve at any flux level previously detected by Swift.  It shows possible fluctuations at most orbital phases, superposed on a broad trend that is consistent with intrabinary shock models.  However, as best illustrated by Figure~\ref{fig:binned} the stochastic nature of the fluctuations makes it impossible to define any underlying ``quiescent'' light curve precisely.  The minimum is evidently close to phase $\phi=0.25$, but the broad maximum may or may not be centered at $\phi=0.75$, the inferior conjunction of the pulsar.  Instead, there may be a plateau that slopes downward from $\phi=0.6$ to $\phi=1$, a common asymmetry in redbacks, e.g., PSR J1023+0038 \citep{bog11}, PSR J2215+5135 \citep{sul24}, and PSR J2129$-$0429 \citep{rom16}.  In any case, such light curves are generally modeled with a shock that is wrapped around the pulsar, with synchrotron emission beamed by particles that flow tangentially to the shock surface \citep{rom16,wad17,wad18,kan19,van20}.  

The low- and high-flux phases in Table~\ref{tab:xspec} have $F_X(0.3-10\,{\rm keV})=4.3\times10^{-13}$ and $6.6\times10^{-13}$ erg~cm$^{-2}$~s$^{-1}$, respectively.  In the context of the Swift monitoring results, these fall within the ``intermediate state'' as categorized in Paper~1.  In comparison, the low flux state in Swift had on average $F_X(0.3-8\,{\rm keV})=1.95\times10^{-13}$ erg~cm$^{-2}$~s$^{-1}$, while the peak of the brightest flare detected by Swift in 2020 had $F_X(0.3-10\,{\rm keV})=1.3\times10^{-11}$ erg~cm$^{-2}$~s$^{-1}$.

There were no counterparts of the X-ray fluctuations in the OM, where the $U$-band light is dominated by the stellar photosphere. This can be understood in that a straight extrapolation of the X-ray power-law spectrum to 3440~\AA, the center of the OM $U$ band, fails to account for even the small excess of OM counts above the photospheric contribution around day 732.5, since it corresponds to $U=24.0$ (before applying interstellar extinction).  In contrast, during the peak of the brightest flares observed by Swift in 2020, the X-ray spectrum, as well the UV to X-ray spectral energy distribution of the brightest flare, appeared to be steeper; see Paper~1 for details.

The phase-averaged X-ray luminosity of $3.3\times10^{32}$~erg~s$^{-1}$ and spectral index $\Gamma=1.53$ are close to the mean values for redbacks: $2.3\times10^{32}$ erg~s$^{-1}$ and $\Gamma=1.44$ \citep{swi22}.  In comparison, the brightest flare previously seen by Swift in 2020 had a peak luminosity of $\approx8\times10^{33}$ erg~s$^{-1}$.  Although the spin-down power $\dot E$ of \obj\ is unknown, the mean ratio $L_x/\dot E$ for redbacks is $\approx2\times10^{-3}$, which would suggest that $\dot E\sim10^{35}$ erg~s$^{-1}$. 

The power-law spectral index is close to 1.5, the smallest that is likely to be produced by diffusive shock acceleration, and shows no change between the opposite phases of the orbit.  Neither is there evidence for thermal plasma emission, either bremsstrahlung or emission lines, in the residuals from the spectral fit.  Similar and even flatter indices in redbacks support the theory that magnetic reconnection accelerates the synchrotron emitting particles \citep{sir11,cor22}, in this case without evidence of reprocessing of energy into hot plasma.  The bright flares in 2020--2021, on the other hand, had a steeper spectrum in the optical through far UV that was not constrained well enough to rule out a thermal contribution, especially in the unobserved extreme UV.

\subsection{Absence of Radio Pulsations}

Many redbacks are difficult to detect as radio pulsars because they are eclipsed during half or more of the orbit and also randomly disappear at any orbital phase (e.g., \citealt{per23}, and references therein).  \obj\ has been searched unsuccessfully for pulsations at all orbital phases \citep{joh25}. Those redbacks and BWs that flare frequently and have strong, variable recombination emission lines are the most elusive, suggesting that these properties, which they share with \obj\, are signatures of dense plasma that can scatter or absorb radio pulses.  One such redback candidate was \fgl\ \citep{hal17a,cho18}, which was only confirmed as \mspone\ \citep{tho24} after several unsuccessful radio campaigns.  It was then possible to bootstrap the entire coherent pulse timing history from Fermi.  Another redback candidate comes from radio imaging of \mspfive\ \citep{zic24}.  It is totally eclipsed during 40\% of its 5.27~hr (presumed) orbital period, implying a radius of the absorbing medium larger than the likely size of a companion star's Roche lobe.  It was not detected as a pulsar even out of eclipse, possibly due to dispersion or scattering by an extended wind.

Among BW and BW candidates, a small fraction are not detected as radio pulsars.  The one BW that is most similar to \obj\ in its behavior is \mspfour, which flares frequently in X-ray and optical/UV \citep{rom12,an17,del22} at all orbital phases and has strong emission lines \citep{rom15}.  In one \xmm\ observation, six X-ray flares recurred with a period of 124~minutes, different from the orbital period of 93.77~minutes.  Its 2.56~ms pulsations were first detected in a blind search of the Fermi $\gamma$-rays \citep{ple12}, and only later in radio \citep{ray13} with the help of the $\gamma$-ray ephemeris.  Even then, only one weak detection lasting $\approx1100$~s was achieved during a total of 22.25~hr of exposure over 20 sessions.  The lesson from these highly enshrouded pulsars is that many repeated observations at different frequencies may eventually encounter a temporary hole through which radio pulses escape, enough to seed a comprehensive Fermi timing solution.

\subsection{Interpretation}

Observations of \obj\ suggest that its companion is the source of a substantial stellar wind that encloses the pulsar with plasma that absorbs any radio pulsations, generates variable recombination emission lines, and emits nonthermal, persistent and fluctuating X-rays when shocked against the pulsar wind.  These phenomena could be explained in terms of the conditions for pressure balance and shock stability investigated by \citet{wad18} in the case of a shock curved around the pulsar, which we recount here.  For a companion wind to be wrapped around the pulsar, it should be gravitationally captured within the Roche lobe of the pulsar, losing angular momentum but not circularizing before reaching the shock; otherwise, an accretion disk would form.  This requires viscosity and heating within the pulsar's Roche lobe, i.e., an advection-dominated accretion flow upstream of the shock.

However, such a shock configuration is unstable on a dynamical (orbital) timescale, requiring some regulating effect, e.g., irradiation of the companion, to stabilize it on the longer timescales it is observed to exist.  This suggested to \citet{wad18} that all redbacks with shocks curved around the pulsar are transitional MSPs (tMSPs), in which the intrabinary shock alternates on timescales of years to decades with a subluminous accretion disk displaying distinctive phenomena that have not yet been seen in \obj; see, e.g., the prototype tMSP PSR J1023+0038 \citep{bog15}.  Variable irradiation feedback on stars that are convection-dominated in their energy transport is likely to induce expansion or contraction of the stellar envelope with an inverse dependence of luminosity on radius and little change in effective temperature, unlike Stefan-Boltzmann scaling \citep{wad18}.  In this interpretation, the decrease in the optical brightness of \obj\ observed over the past 10 years indicates an increasing Roche-lobe filling fraction that may allow the mass-loss rate or circularization radius to increase, predicting an eventual transition to the accretion-disk state.

\section{Conclusions}
\label{sec:conc}

The first X-ray observation to study the highly variable candidate redback \obj\ over its entire 16.5~hr orbit finds it at a flux $\approx2-3$ times greater than the poorly characterized minimum that Swift measured previously.  The X-ray spectrum is consistent with a power law of $\Gamma=1.53$, independent of the orbital phase, and similar to that of other redbacks.  Fluctuations around a smooth light curve make it difficult to study the quiescent structure of the underlying intrabinary shock, if it even has a stable configuration emitting at half the present flux or less.  Because the broad minimum in the flux is centered at $\phi=0.25$, the inferior conjunction of the companion as determined by prior optical spectroscopy and continued photometry, favored models would have the shock wrapped around the pulsar.  The simultaneous $U$-band light curve from the OM is dominated by ellipsoidal modulation of the photospheric emission of the bright companion star, consistent with previous ground-based optical and Swift photometry.

The failure to detect radio pulsations from \obj\ may be related to the unusually massive companion (for a redback). Its Roche lobe presents a large target for the pulsar wind, which might help to generate a strong stellar wind capable of enshrouding the pulsar and absorbing or scattering radio pulses.

\begin{acknowledgements}
  This investigation is based on an observation made by \xmm, an ESA science mission with instruments and contributions directly funded by ESA Member States and NASA, and includes data obtained at the MDM Observatory, operated by Dartmouth College, Columbia University, The Ohio State University, Ohio University, and the University of Michigan.  Support for this work was provided by NASA \xmm\ guest observer grant 80NSSC24K1517 and Fermi grant 80NSSC26K0304.  The Zwicky Transient Facility is supported by the National Science Foundation under Grants No. AST-1440341 and AST-2034437 and a collaboration including current partners Caltech, IPAC, the Oskar Klein Center at Stockholm University, the University of Maryland, University of California, Berkeley, the University of Wisconsin at Milwaukee, the University of Warwick, Ruhr University, Cornell University, Northwestern University and Drexel University. Operations are conducted by COO, IPAC, and UW. The ZTF data are available at \dataset[doi:10.26131/IRSA598]{https://doi.org/10.26131/irsa598}. The ATLAS project is primarily funded to search for near earth asteroids through NASA grants NN12AR55G, 80NSSC18K0284, and 80NSSC18K1575. The ATLAS science products are made possible through the contributions of the University of Hawaii Institute for Astronomy, The Queen's University Belfast, the Space Telescope Science Institute, the South African Astronomical Observatory, and The Millennium Institute of Astrophysics (MAS), Chile.
  The XMM-Newton Science Analysis Software (SAS) is developed and maintained by the Science Operations Centre at the European Space Astronomy Centre and the Survey Science Centre at the University of Leicester. This research has made use of data and software provided by the High Energy Astrophysics Science Archive Research Center (HEASARC), which is a service of the Astrophysics Science Division at NASA/GSFC. We acknowledge use of NASA's Astrophysics Data System (ADS) Bibliographic Services and the arXiv.
\end{acknowledgements}

\facilities{ATLAS, McGraw-Hill, XMM}

\software{HEASoft \citep{2014ascl.soft08004N}, SAS \citep{2004ASPC..314..759G}}

\bibliography{ms.bib}

@ARTICLE{aln21,
       author = {{Al Noori}, Hind and {Roberts}, Mallory S.~E. and {Torres}, Rodrigo A. and {McLaughlin}, Maura A. and {Gentile}, Peter A. and {Hessels}, Jason W.~T. and {Ray}, Paul S. and {Kerr}, Matthew and {Breton}, Rene P.},
        title = "{X-Ray and Optical Studies of the Redback System PSR J2129-0429}",
      journal = {\apj},
         year = 2018,
        month = jul,
       volume = {861},
       number = {2},
          eid = {89},
        pages = {89},
          doi = {10.3847/1538-4357/aac828},
       adsurl = {https://ui.adsabs.harvard.edu/abs/2018ApJ...861...89A}
}

@ARTICLE{alp82,
       author = {{Alpar}, M.~A. and {Cheng}, A.~F. and {Ruderman}, M.~A. and {Shaham}, J.},
        title = "{A new class of radio pulsars}",
      journal = {\nat},
         year = 1982,
        month = dec,
       volume = {300},
       number = {5894},
        pages = {728-730},
          doi = {10.1038/300728a0},
       adsurl = {https://ui.adsabs.harvard.edu/abs/1982Natur.300..728A}
}

@ARTICLE{an17,
       author = {{An}, Hongjun and {Romani}, Roger W. and {Johnson}, Tyrel and {Kerr}, Matthew and {Clark}, Colin J.},
        title = "{High-energy Variability of PSR J1311-3430}",
      journal = {\apj},
         year = 2017,
        month = nov,
       volume = {850},
       number = {1},
          eid = {100},
        pages = {100},
          doi = {10.3847/1538-4357/aa947f},
archivePrefix = {arXiv},
       eprint = {1710.06097},
 primaryClass = {astro-ph.HE},
       adsurl = {https://ui.adsabs.harvard.edu/abs/2017ApJ...850..100A}
}

@ARTICLE{bel16,
       author = {{Bellm}, Eric C. and {Kaplan}, David L. and {Breton}, Rene P. and {Phinney}, E. Sterl and {Bhalerao}, Varun B. and {Camilo}, Fernando and {Dahal}, Sumit and {Djorgovski}, S.~G. and {Drake}, Andrew J. and {Hessels}, J.~W.~T. and {Laher}, Russ R. and {Levitan}, David B. and {Lewis}, Fraser and {Mahabal}, Ashish A. and {Ofek}, Eran O. and {Prince}, Thomas A. and {Ransom}, Scott M. and {Roberts}, Mallory S.~E. and {Russell}, David M. and {Sesar}, Branimir and {Surace}, Jason A. and {Tang}, Sumin},
        title = "{Properties and Evolution of the Redback Millisecond Pulsar Binary PSR J2129-0429}",
      journal = {\apj},
         year = 2016,
        month = jan,
       volume = {816},
       number = {2},
          eid = {74},
        pages = {74},
          doi = {10.3847/0004-637X/816/2/74},
archivePrefix = {arXiv},
       eprint = {1510.00721},
 primaryClass = {astro-ph.SR},
       adsurl = {https://ui.adsabs.harvard.edu/abs/2016ApJ...816...74B}
}

@ARTICLE{bel19,
       author = {{Bellm}, Eric C. and {Kulkarni}, Shrinivas R. and {Graham}, Matthew J. and {Dekany}, Richard and {Smith}, Roger M. and {Riddle}, Reed and {Masci}, Frank J. and {Helou}, George and {Prince}, Thomas A. and {Adams}, Scott M. and {Barbarino}, C. and {Barlow}, Tom and {Bauer}, James and {Beck}, Ron and {Belicki}, Justin and {Biswas}, Rahul and {Blagorodnova}, Nadejda and {Bodewits}, Dennis and {Bolin}, Bryce and {Brinnel}, Valery and {Brooke}, Tim and {Bue}, Brian and {Bulla}, Mattia and {Burruss}, Rick and {Cenko}, S. Bradley and {Chang}, Chan-Kao and {Connolly}, Andrew and {Coughlin}, Michael and {Cromer}, John and {Cunningham}, Virginia and {De}, Kishalay and {Delacroix}, Alex and {Desai}, Vandana and {Duev}, Dmitry A. and {Eadie}, Gwendolyn and {Farnham}, Tony L. and {Feeney}, Michael and {Feindt}, Ulrich and {Flynn}, David and {Franckowiak}, Anna and {Frederick}, S. and {Fremling}, C. and {Gal-Yam}, Avishay and {Gezari}, Suvi and {Giomi}, Matteo and {Goldstein}, Daniel A. and {Golkhou}, V. Zach and {Goobar}, Ariel and {Groom}, Steven and {Hacopians}, Eugean and {Hale}, David and {Henning}, John and {Ho}, Anna Y.~Q. and {Hover}, David and {Howell}, Justin and {Hung}, Tiara and {Huppenkothen}, Daniela and {Imel}, David and {Ip}, Wing-Huen and {Ivezi{\'c}}, {\v{Z}}eljko and {Jackson}, Edward and {Jones}, Lynne and {Juric}, Mario and {Kasliwal}, Mansi M. and {Kaspi}, S. and {Kaye}, Stephen and {Kelley}, Michael S.~P. and {Kowalski}, Marek and {Kramer}, Emily and {Kupfer}, Thomas and {Landry}, Walter and {Laher}, Russ R. and {Lee}, Chien-De and {Lin}, Hsing Wen and {Lin}, Zhong-Yi and {Lunnan}, Ragnhild and {Giomi}, Matteo and {Mahabal}, Ashish and {Mao}, Peter and {Miller}, Adam A. and {Monkewitz}, Serge and {Murphy}, Patrick and {Ngeow}, Chow-Choong and {Nordin}, Jakob and {Nugent}, Peter and {Ofek}, Eran and {Patterson}, Maria T. and {Penprase}, Bryan and {Porter}, Michael and {Rauch}, Ludwig and {Rebbapragada}, Umaa and {Reiley}, Dan and {Rigault}, Mickael and {Rodriguez}, Hector and {van Roestel}, Jan and {Rusholme}, Ben and {van Santen}, Jakob and {Schulze}, S. and {Shupe}, David L. and {Singer}, Leo P. and {Soumagnac}, Maayane T. and {Stein}, Robert and {Surace}, Jason and {Sollerman}, Jesper and {Szkody}, Paula and {Taddia}, F. and {Terek}, Scott and {Van Sistine}, Angela and {van Velzen}, Sjoert and {Vestrand}, W. Thomas and {Walters}, Richard and {Ward}, Charlotte and {Ye}, Quan-Zhi and {Yu}, Po-Chieh and {Yan}, Lin and {Zolkower}, Jeffry},
        title = "{The Zwicky Transient Facility: System Overview, Performance, and First Results}",
      journal = {\pasp},
         year = 2019,
        month = jan,
       volume = {131},
       number = {995},
        pages = {018002},
          doi = {10.1088/1538-3873/aaecbe},
archivePrefix = {arXiv},
       eprint = {1902.01932},
 primaryClass = {astro-ph.IM},
       adsurl = {https://ui.adsabs.harvard.edu/abs/2019PASP..131a8002B}
}

@ARTICLE{bog15,
       author = {{Bogdanov}, Slavko and {Archibald}, Anne M. and {Bassa}, Cees and {Deller}, Adam T. and {Halpern}, Jules P. and {Heald}, George and {Hessels}, Jason W.~T. and {Janssen}, Gemma H. and {Lyne}, Andrew G. and {Mold{\'o}n}, Javier and {Paragi}, Zsolt and {Patruno}, Alessandro and {Perera}, Benetge B.~P. and {Stappers}, Ben W. and {Tendulkar}, Shriharsh P. and {D'Angelo}, Caroline R. and {Wijnands}, Rudy},
        title = "{Coordinated X-Ray, Ultraviolet, Optical, and Radio Observations of the PSR J1023+0038 System in a Low-mass X-Ray Binary State}",
      journal = {\apj},
         year = 2015,
        month = jun,
       volume = {806},
       number = {2},
          eid = {148},
        pages = {148},
          doi = {10.1088/0004-637X/806/2/148},
archivePrefix = {arXiv},
       eprint = {1412.5145},
 primaryClass = {astro-ph.HE},
       adsurl = {https://ui.adsabs.harvard.edu/abs/2015ApJ...806..148B}
}

@ARTICLE{bog05,
       author = {{Bogdanov}, Slavko and {Grindlay}, Jonathan E. and {van den Berg}, Maureen},
        title = "{An X-Ray Variable Millisecond Pulsar in the Globular Cluster 47 Tucanae: Closing the Link to Low-Mass X-Ray Binaries}",
      journal = {\apj},
         year = 2005,
        month = sep,
       volume = {630},
       number = {2},
        pages = {1029-1036},
          doi = {10.1086/432249},
archivePrefix = {arXiv},
       eprint = {astro-ph/0506031},
 primaryClass = {astro-ph},
       adsurl = {https://ui.adsabs.harvard.edu/abs/2005ApJ...630.1029B}
}

@ARTICLE{bog11,
       author = {{Bogdanov}, Slavko and {Archibald}, Anne M. and {Hessels}, Jason W.~T. and {Kaspi}, Victoria M. and {Lorimer}, Duncan and {McLaughlin}, Maura A. and {Ransom}, Scott M. and {Stairs}, Ingrid H.},
        title = "{A Chandra X-Ray Observation of the Binary Millisecond Pulsar PSR J1023+0038}",
      journal = {\apj},
         year = 2011,
        month = dec,
       volume = {742},
       number = {2},
          eid = {97},
        pages = {97},
          doi = {10.1088/0004-637X/742/2/97},
archivePrefix = {arXiv},
       eprint = {1108.5753},
 primaryClass = {astro-ph.HE},
       adsurl = {https://ui.adsabs.harvard.edu/abs/2011ApJ...742...97B}
}

@ARTICLE{bog14,
       author = {{Bogdanov}, Slavko and {Patruno}, Alessandro and {Archibald}, Anne M. and {Bassa}, Cees and {Hessels}, Jason W.~T. and {Janssen}, Gemma H. and {Stappers}, Ben W.},
        title = "{X-Ray Observations of XSS J12270-4859 in a New Low State: A Transformation to a Disk-free Rotation-powered Pulsar Binary}",
      journal = {\apj},
         year = 2014,
        month = jul,
       volume = {789},
       number = {1},
          eid = {40},
        pages = {40},
          doi = {10.1088/0004-637X/789/1/40},
archivePrefix = {arXiv},
       eprint = {1402.6324},
 primaryClass = {astro-ph.HE},
       adsurl = {https://ui.adsabs.harvard.edu/abs/2014ApJ...789...40B}
}

@ARTICLE{bro21,
       author = {{Gaia Collaboration} and {Brown}, A.~G.~A. and {Vallenari}, A. and {Prusti}, T. and {de Bruijne}, J.~H.~J. and {Babusiaux}, C. and {Biermann}, M. and {Creevey}, O.~L. and {Evans}, D.~W. and {Eyer}, L. and {Hutton}, A. and {Jansen}, F. and {Jordi}, C. and {Klioner}, S.~A. and {Lammers}, U. and {Lindegren}, L. and {Luri}, X. and {Mignard}, F. and {Panem}, C. and {Pourbaix}, D. and {Randich}, S. and {Sartoretti}, P. and {Soubiran}, C. and {Walton}, N.~A. and {Arenou}, F. and {Bailer-Jones}, C.~A.~L. and {Bastian}, U. and {Cropper}, M. and {Drimmel}, R. and {Katz}, D. and {Lattanzi}, M.~G. and {van Leeuwen}, F. and {Bakker}, J. and {Cacciari}, C. and {Casta{\~n}eda}, J. and {De Angeli}, F. and {Ducourant}, C. and {Fabricius}, C. and {Fouesneau}, M. and {Fr{\'e}mat}, Y. and {Guerra}, R. and {Guerrier}, A. and {Guiraud}, J. and {Jean-Antoine Piccolo}, A. and {Masana}, E. and {Messineo}, R. and {Mowlavi}, N. and {Nicolas}, C. and {Nienartowicz}, K. and {Pailler}, F. and {Panuzzo}, P. and {Riclet}, F. and {Roux}, W. and {Seabroke}, G.~M. and {Sordo}, R. and {Tanga}, P. and {Th{\'e}venin}, F. and {Gracia-Abril}, G. and {Portell}, J. and {Teyssier}, D. and {Altmann}, M. and {Andrae}, R. and {Bellas-Velidis}, I. and {Benson}, K. and {Berthier}, J. and {Blomme}, R. and {Brugaletta}, E. and {Burgess}, P.~W. and {Busso}, G. and {Carry}, B. and {Cellino}, A. and {Cheek}, N. and {Clementini}, G. and {Damerdji}, Y. and {Davidson}, M. and {Delchambre}, L. and {Dell'Oro}, A. and {Fern{\'a}ndez-Hern{\'a}ndez}, J. and {Galluccio}, L. and {Garc{\'\i}a-Lario}, P. and {Garcia-Reinaldos}, M. and {Gonz{\'a}lez-N{\'u}{\~n}ez}, J. and {Gosset}, E. and {Haigron}, R. and {Halbwachs}, J.-L. and {Hambly}, N.~C. and {Harrison}, D.~L. and {Hatzidimitriou}, D. and {Heiter}, U. and {Hern{\'a}ndez}, J. and {Hestroffer}, D. and {Hodgkin}, S.~T. and {Holl}, B. and {Jan{\ss}en}, K. and {Jevardat de Fombelle}, G. and {Jordan}, S. and {Krone-Martins}, A. and {Lanzafame}, A.~C. and {L{\"o}ffler}, W. and {Lorca}, A. and {Manteiga}, M. and {Marchal}, O. and {Marrese}, P.~M. and {Moitinho}, A. and {Mora}, A. and {Muinonen}, K. and {Osborne}, P. and {Pancino}, E. and {Pauwels}, T. and {Petit}, J.-M. and {Recio-Blanco}, A. and {Richards}, P.~J. and {Riello}, M. and {Rimoldini}, L. and {Robin}, A.~C. and {Roegiers}, T. and {Rybizki}, J. and {Sarro}, L.~M. and {Siopis}, C. and {Smith}, M. and {Sozzetti}, A. and {Ulla}, A. and {Utrilla}, E. and {van Leeuwen}, M. and {van Reeven}, W. and {Abbas}, U. and {Abreu Aramburu}, A. and {Accart}, S. and {Aerts}, C. and {Aguado}, J.~J. and {Ajaj}, M. and {Altavilla}, G. and {{\'A}lvarez}, M.~A. and {{\'A}lvarez Cid-Fuentes}, J. and {Alves}, J. and {Anderson}, R.~I. and {Anglada Varela}, E. and {Antoja}, T. and {Audard}, M. and {Baines}, D. and {Baker}, S.~G. and {Balaguer-N{\'u}{\~n}ez}, L. and {Balbinot}, E. and {Balog}, Z. and {Barache}, C. and {Barbato}, D. and {Barros}, M. and {Barstow}, M.~A. and {Bartolom{\'e}}, S. and {Bassilana}, J.-L. and {Bauchet}, N. and {Baudesson-Stella}, A. and {Becciani}, U. and {Bellazzini}, M. and {Bernet}, M. and {Bertone}, S. and {Bianchi}, L. and {Blanco-Cuaresma}, S. and {Boch}, T. and {Bombrun}, A. and {Bossini}, D. and {Bouquillon}, S. and {Bragaglia}, A. and {Bramante}, L. and {Breedt}, E. and {Bressan}, A. and {Brouillet}, N. and {Bucciarelli}, B. and {Burlacu}, A. and {Busonero}, D. and {Butkevich}, A.~G. and {Buzzi}, R. and {Caffau}, E. and {Cancelliere}, R. and {C{\'a}novas}, H. and {Cantat-Gaudin}, T. and {Carballo}, R. and {Carlucci}, T. and {Carnerero}, M.~I. and {Carrasco}, J.~M. and {Casamiquela}, L. and {Castellani}, M. and {Castro-Ginard}, A. and {Castro Sampol}, P. and {Chaoul}, L. and {Charlot}, P. and {Chemin}, L. and {Chiavassa}, A. and {Cioni}, M.-R.~L. and {Comoretto}, G. and {Cooper}, W.~J. and {Cornez}, T. and {Cowell}, S. and {Crifo}, F. and {Crosta}, M. and {Crowley}, C. and {Dafonte}, C. and {Dapergolas}, A. and {David}, M. and {David}, P.},
        title = "{Gaia Early Data Release 3. Summary of the contents and survey properties}",
      journal = {\aap},
         year = 2021,
        month = may,
       volume = {649},
          eid = {A1},
        pages = {A1},
          doi = {10.1051/0004-6361/202039657},
archivePrefix = {arXiv},
       eprint = {2012.01533},
 primaryClass = {astro-ph.GA},
       adsurl = {https://ui.adsabs.harvard.edu/abs/2021A&A...649A...1G}
}

@ARTICLE{cho18,
       author = {{Cho}, Patricia B. and {Halpern}, Jules P. and {Bogdanov}, Slavko},
        title = "{Variable Heating and Flaring of Three Redback Millisecond Pulsar Companions}",
      journal = {\apj},
         year = 2018,
        month = oct,
       volume = {866},
       number = {1},
          eid = {71},
        pages = {71},
          doi = {10.3847/1538-4357/aade92},
archivePrefix = {arXiv},
       eprint = {1809.00215},
 primaryClass = {astro-ph.HE},
       adsurl = {https://ui.adsabs.harvard.edu/abs/2018ApJ...866...71C}
}

@ARTICLE{cla11,
       author = {{Claret}, A. and {Bloemen}, S.},
        title = "{Gravity and limb-darkening coefficients for the Kepler, CoRoT, Spitzer, uvby, UBVRIJHK, and Sloan photometric systems}",
      journal = {\aap},
         year = 2011,
        month = may,
       volume = {529},
          eid = {A75},
        pages = {A75},
          doi = {10.1051/0004-6361/201116451},
       adsurl = {https://ui.adsabs.harvard.edu/abs/2011A&A...529A..75C}
}

@ARTICLE{cor22,
       author = {{Cort{\'e}s}, Jorge and {Sironi}, Lorenzo},
        title = "{Global Kinetic Modeling of the Intrabinary Shock in Spider Pulsars}",
      journal = {\apj},
         year = 2022,
        month = jul,
       volume = {933},
       number = {2},
          eid = {140},
        pages = {140},
          doi = {10.3847/1538-4357/ac74b2},
archivePrefix = {arXiv},
       eprint = {2203.00023},
 primaryClass = {astro-ph.HE},
       adsurl = {https://ui.adsabs.harvard.edu/abs/2022ApJ...933..140C}
}

@ARTICLE{cor24,
       author = {{Cort{\'e}s}, Jorge and {Sironi}, Lorenzo},
        title = "{Particle acceleration and non-thermal emission at the intrabinary shock of spider pulsars - I. Non-radiative simulations}",
      journal = {\mnras},
         year = 2024,
        month = nov,
       volume = {534},
       number = {3},
        pages = {2551-2565},
          doi = {10.1093/mnras/stae2278},
archivePrefix = {arXiv},
       eprint = {2404.03700},
 primaryClass = {astro-ph.HE},
       adsurl = {https://ui.adsabs.harvard.edu/abs/2024MNRAS.534.2551C}
}

@ARTICLE{cor25,
       author = {{Cort{\'e}s}, Jorge and {Sironi}, Lorenzo},
        title = "{Particle acceleration and non-thermal emission at the intrabinary shock of spider pulsars ─ II. Fast-cooling simulations}",
      journal = {\mnras},
         year = 2025,
        month = sep,
       volume = {542},
       number = {2},
        pages = {917-926},
          doi = {10.1093/mnras/staf284},
archivePrefix = {arXiv},
       eprint = {2501.08387},
 primaryClass = {astro-ph.HE},
       adsurl = {https://ui.adsabs.harvard.edu/abs/2025MNRAS.542..917C}
}

@ARTICLE{del22,
       author = {{De Luca}, Andrea and {Marelli}, Martino and {Mereghetti}, Sandro and {Salvaterra}, Ruben and {Mignani}, Roberto and {Belfiore}, Andrea},
        title = "{A puzzling 2-hour X-ray periodicity in the 1.5-hour orbital period black widow PSR J1311‒3430}",
      journal = {\aap},
         year = 2022,
        month = nov,
       volume = {667},
          eid = {L7},
        pages = {L7},
          doi = {10.1051/0004-6361/202244643},
archivePrefix = {arXiv},
       eprint = {2210.10806},
 primaryClass = {astro-ph.HE},
       adsurl = {https://ui.adsabs.harvard.edu/abs/2022A&A...667L...7D}
}

@ARTICLE{den16,
       author = {{Deneva}, J.~S. and {Ray}, P.~S. and {Camilo}, F. and {Halpern}, J.~P. and {Wood}, K. and {Cromartie}, H.~T. and {Ferrara}, E. and {Kerr}, M. and {Ransom}, S.~M. and {Wolff}, M.~T. and {Chambers}, K.~C. and {Magnier}, E.~A.},
        title = "{Multiwavelength Observations of the Redback Millisecond Pulsar J1048+2339}",
      journal = {\apj},
         year = 2016,
        month = jun,
       volume = {823},
       number = {2},
          eid = {105},
        pages = {105},
          doi = {10.3847/0004-637X/823/2/105},
archivePrefix = {arXiv},
       eprint = {1601.03681},
 primaryClass = {astro-ph.HE},
       adsurl = {https://ui.adsabs.harvard.edu/abs/2016ApJ...823..105D}
}

@ARTICLE{den21,
       author = {{Deneva}, J.~S. and {Ray}, P.~S. and {Camilo}, F. and {Freire}, P.~C.~C. and {Cromartie}, H.~T. and {Ransom}, S.~M. and {Ferrara}, E. and {Kerr}, M. and {Burnett}, T.~H. and {Parkinson}, P.~M. Saz},
        title = "{Timing of Eight Binary Millisecond Pulsars Found with Arecibo in Fermi-LAT Unidentified Sources}",
      journal = {\apj},
         year = 2021,
        month = mar,
       volume = {909},
       number = {1},
          eid = {6},
        pages = {6},
          doi = {10.3847/1538-4357/abd7a1},
archivePrefix = {arXiv},
       eprint = {2012.15185},
 primaryClass = {astro-ph.HE},
       adsurl = {https://ui.adsabs.harvard.edu/abs/2021ApJ...909....6D}
}

@ARTICLE{dra09,
       author = {{Drake}, A.~J. and {Djorgovski}, S.~G. and {Mahabal}, A. and {Beshore}, E. and {Larson}, S. and {Graham}, M.~J. and {Williams}, R. and {Christensen}, E. and {Catelan}, M. and {Boattini}, A. and {Gibbs}, A. and {Hill}, R. and {Kowalski}, R.},
        title = "{First Results from the Catalina Real-Time Transient Survey}",
      journal = {\apj},
         year = 2009,
        month = may,
       volume = {696},
       number = {1},
        pages = {870-884},
          doi = {10.1088/0004-637X/696/1/870},
archivePrefix = {arXiv},
       eprint = {0809.1394},
 primaryClass = {astro-ph},
       adsurl = {https://ui.adsabs.harvard.edu/abs/2009ApJ...696..870D}
}

@ARTICLE{eas10,
       author = {{Eastman}, Jason and {Siverd}, Robert and {Gaudi}, B. Scott},
        title = "{Achieving Better Than 1 Minute Accuracy in the Heliocentric and Barycentric Julian Dates}",
      journal = {\pasp},
         year = 2010,
        month = aug,
       volume = {122},
       number = {894},
        pages = {935},
          doi = {10.1086/655938},
archivePrefix = {arXiv},
       eprint = {1005.4415},
 primaryClass = {astro-ph.IM},
       adsurl = {https://ui.adsabs.harvard.edu/abs/2010PASP..122..935E}
}

@ARTICLE{gom21,
       author = {{Gomel}, Roy and {Faigler}, Simchon and {Mazeh}, Tsevi},
        title = "{Search for dormant black holes in ellipsoidal variables I. Revisiting the expected amplitudes of the photometric modulation}",
      journal = {\mnras},
         year = 2021,
        month = feb,
       volume = {501},
       number = {2},
        pages = {2822-2832},
          doi = {10.1093/mnras/staa3305},
archivePrefix = {arXiv},
       eprint = {2008.11209},
 primaryClass = {astro-ph.SR},
       adsurl = {https://ui.adsabs.harvard.edu/abs/2021MNRAS.501.2822G}
}

@ARTICLE{hal17,
       author = {{Halpern}, Jules P. and {Bogdanov}, Slavko and {Thorstensen}, John R.},
        title = "{X-Ray and Optical Study of the Gamma-ray Source 3FGL J0838.8-2829: Identification of a Candidate Millisecond Pulsar Binary and an Asynchronous Polar}",
      journal = {\apj},
         year = 2017,
        month = apr,
       volume = {838},
       number = {2},
          eid = {124},
        pages = {124},
          doi = {10.3847/1538-4357/838/2/124},
archivePrefix = {arXiv},
       eprint = {1701.04116},
 primaryClass = {astro-ph.HE},
       adsurl = {https://ui.adsabs.harvard.edu/abs/2017ApJ...838..124H}
}

@ARTICLE{hal22,
       author = {{Halpern}, Jules P. and {Perez}, Karen I. and {Bogdanov}, Slavko},
        title = "{Luminous Optical and X-Ray Flaring of the Putative Redback Millisecond Pulsar 1FGL J0523.5-2529}",
      journal = {\apj},
         year = 2022,
        month = aug,
       volume = {935},
       number = {2},
          eid = {151},
        pages = {151},
          doi = {10.3847/1538-4357/ac8161},
archivePrefix = {arXiv},
       eprint = {2207.08198},
 primaryClass = {astro-ph.HE},
       adsurl = {https://ui.adsabs.harvard.edu/abs/2022ApJ...935..151H}
}

@ARTICLE{hal17a,
       author = {{Halpern}, J.~P. and {Strader}, J. and {Li}, M.},
        title = "{A Likely Redback Millisecond Pulsar Counterpart of 3FGL J0838.8-2829}",
      journal = {\apj},
         year = 2017,
        month = aug,
       volume = {844},
       number = {2},
          eid = {150},
        pages = {150},
          doi = {10.3847/1538-4357/aa7cff},
archivePrefix = {arXiv},
       eprint = {1706.09511},
 primaryClass = {astro-ph.HE},
       adsurl = {https://ui.adsabs.harvard.edu/abs/2017ApJ...844..150H}
}

@ARTICLE{joh25,
       author = {{Johnson}, Owen. A. and {Keane}, E.~F. and {McKenna}, D.~J. and {Qiu}, H. and {Swihart}, S.~J. and {Strader}, J. and {McLaughlin}, M.},
        title = "{Radio Observations of a Candidate Redback Millisecond Pulsar: 1FGL J0523.5-2529}",
      journal = {The Open Journal of Astrophysics},
         year = 2025,
        month = nov,
       volume = {8},
        pages = {47516},
          doi = {10.33232/001c.147516},
archivePrefix = {arXiv},
       eprint = {2508.15435},
 primaryClass = {astro-ph.HE},
       adsurl = {https://ui.adsabs.harvard.edu/abs/2025OJAp....847516J}
}

@ARTICLE{kan19,
       author = {{Kandel}, D. and {Romani}, Roger W. and {An}, Hongjun},
        title = "{The Synchrotron Emission Pattern of Intrabinary Shocks}",
      journal = {\apj},
         year = 2019,
        month = jul,
       volume = {879},
       number = {2},
          eid = {73},
        pages = {73},
          doi = {10.3847/1538-4357/ab24d9},
archivePrefix = {arXiv},
       eprint = {1905.12591},
 primaryClass = {astro-ph.HE},
       adsurl = {https://ui.adsabs.harvard.edu/abs/2019ApJ...879...73K}
}

@ARTICLE{kol25,
       author = {{Koljonen}, Karri I.~I. and {Linares}, Manuel},
        title = "{SpiderCat: A Catalog of Compact Binary Millisecond Pulsars}",
      journal = {\apj},
         year = 2025,
        month = nov,
       volume = {994},
       number = {1},
          eid = {8},
        pages = {8},
          doi = {10.3847/1538-4357/ae08a5},
archivePrefix = {arXiv},
       eprint = {2505.11691},
 primaryClass = {astro-ph.HE},
       adsurl = {https://ui.adsabs.harvard.edu/abs/2025ApJ...994....8K}
}

@ARTICLE{li14,
       author = {{Li}, Miao and {Halpern}, Jules P. and {Thorstensen}, John R.},
        title = "{Optical Counterparts of Two Fermi Millisecond Pulsars: PSR J1301+0833 and PSR J1628-3205}",
      journal = {\apj},
         year = 2014,
        month = nov,
       volume = {795},
       number = {2},
          eid = {115},
        pages = {115},
          doi = {10.1088/0004-637X/795/2/115},
archivePrefix = {arXiv},
       eprint = {1409.3877},
 primaryClass = {astro-ph.HE},
       adsurl = {https://ui.adsabs.harvard.edu/abs/2014ApJ...795..115L}
}

@ARTICLE{mor93,
       author = {{Morris}, Steven L. and {Naftilan}, Stephen A.},
        title = "{The Equations of Ellipsoidal Star Variability Applied to HR 8427}",
      journal = {\apj},
         year = 1993,
        month = dec,
       volume = {419},
        pages = {344},
          doi = {10.1086/173488},
       adsurl = {https://ui.adsabs.harvard.edu/abs/1993ApJ...419..344M}
}

@ARTICLE{per23,
       author = {{Perez}, Karen I. and {Bogdanov}, Slavko and {Halpern}, Jules P. and {Gajjar}, Vishal},
        title = "{Green Bank Telescope Discovery of the Redback Binary Millisecond Pulsar PSR J0212+5321}",
      journal = {\apj},
         year = 2023,
        month = aug,
       volume = {952},
       number = {2},
          eid = {150},
        pages = {150},
          doi = {10.3847/1538-4357/acdc23},
archivePrefix = {arXiv},
       eprint = {2306.04951},
 primaryClass = {astro-ph.HE},
       adsurl = {https://ui.adsabs.harvard.edu/abs/2023ApJ...952..150P}
}

@ARTICLE{ple12,
       author = {{Pletsch}, H.~J. and {Guillemot}, L. and {Fehrmann}, H. and {Allen}, B. and {Kramer}, M. and {Aulbert}, C. and {Ackermann}, M. and {Ajello}, M. and {de Angelis}, A. and {Atwood}, W.~B. and {Baldini}, L. and {Ballet}, J. and {Barbiellini}, G. and {Bastieri}, D. and {Bechtol}, K. and {Bellazzini}, R. and {Borgland}, A.~W. and {Bottacini}, E. and {Brandt}, T.~J. and {Bregeon}, J. and {Brigida}, M. and {Bruel}, P. and {Buehler}, R. and {Buson}, S. and {Caliandro}, G.~A. and {Cameron}, R.~A. and {Caraveo}, P.~A. and {Casandjian}, J.~M. and {Cecchi}, C. and {{\c{C}}elik}, {\"O}. and {Charles}, E. and {Chaves}, R.~C.~G. and {Cheung}, C.~C. and {Chiang}, J. and {Ciprini}, S. and {Claus}, R. and {Cohen-Tanugi}, J. and {Conrad}, J. and {Cutini}, S. and {D'Ammando}, F. and {Dermer}, C.~D. and {Digel}, S.~W. and {Drell}, P.~S. and {Drlica-Wagner}, A. and {Dubois}, R. and {Dumora}, D. and {Favuzzi}, C. and {Ferrara}, E.~C. and {Franckowiak}, A. and {Fukazawa}, Y. and {Fusco}, P. and {Gargano}, F. and {Gehrels}, N. and {Germani}, S. and {Giglietto}, N. and {Giordano}, F. and {Giroletti}, M. and {Godfrey}, G. and {Grenier}, I.~A. and {Grondin}, M.-H. and {Grove}, J.~E. and {Guiriec}, S. and {Hadasch}, D. and {Hanabata}, Y. and {Harding}, A.~K. and {den Hartog}, P.~R. and {Hayashida}, M. and {Hays}, E. and {Hill}, A.~B. and {Hou}, X. and {Hughes}, R.~E. and {J{\'o}hannesson}, G. and {Jackson}, M.~S. and {Jogler}, T. and {Johnson}, A.~S. and {Johnson}, W.~N. and {Kataoka}, J. and {Kerr}, M. and {Kn{\"o}dlseder}, J. and {Kuss}, M. and {Lande}, J. and {Larsson}, S. and {Latronico}, L. and {Lemoine-Goumard}, M. and {Longo}, F. and {Loparco}, F. and {Lovellette}, M.~N. and {Lubrano}, P. and {Massaro}, F. and {Mayer}, M. and {Mazziotta}, M.~N. and {McEnery}, J.~E. and {Mehault}, J. and {Michelson}, P.~F. and {Mitthumsiri}, W. and {Mizuno}, T. and {Monzani}, M.~E. and {Morselli}, A. and {Moskalenko}, I.~V. and {Murgia}, S. and {Nakamori}, T. and {Nemmen}, R. and {Nuss}, E. and {Ohno}, M. and {Ohsugi}, T. and {Omodei}, N. and {Orienti}, M. and {Orlando}, E. and {de Palma}, F. and {Paneque}, D. and {Perkins}, J.~S. and {Piron}, F. and {Pivato}, G. and {Porter}, T.~A. and {Rain{\`o}}, S. and {Rando}, R. and {Ray}, P.~S. and {Razzano}, M. and {Reimer}, A. and {Reimer}, O. and {Reposeur}, T. and {Ritz}, S. and {Romani}, R.~W. and {Romoli}, C. and {Sanchez}, D.~A. and {Parkinson}, P.~M. Saz and {Schulz}, A. and {Sgr{\`o}}, C. and {do Couto e Silva}, E. and {Siskind}, E.~J. and {Smith}, D.~A. and {Spandre}, G. and {Spinelli}, P. and {Suson}, D.~J. and {Takahashi}, H. and {Tanaka}, T. and {Thayer}, J.~B. and {Thayer}, J.~G. and {Thompson}, D.~J. and {Tibaldo}, L. and {Tinivella}, M. and {Troja}, E. and {Usher}, T.~L. and {Vandenbroucke}, J. and {Vasileiou}, V. and {Vianello}, G. and {Vitale}, V. and {Waite}, A.~P. and {Winer}, B.~L. and {Wood}, K.~S. and {Wood}, M. and {Yang}, Z. and {Zimmer}, S.},
        title = "{Binary Millisecond Pulsar Discovery via Gamma-Ray Pulsations}",
      journal = {Science},
         year = 2012,
        month = dec,
       volume = {338},
       number = {6112},
        pages = {1314},
          doi = {10.1126/science.1229054},
archivePrefix = {arXiv},
       eprint = {1211.1385},
 primaryClass = {astro-ph.HE},
       adsurl = {https://ui.adsabs.harvard.edu/abs/2012Sci...338.1314P}
}

@ARTICLE{rad82,
       author = {{Radhakrishnan}, V. and {Srinivasan}, G.},
        title = "{On the origin of the recently discovered ultra-rapid pulsar}",
      journal = {Current Science},
         year = 1982,
        month = dec,
       volume = {51},
        pages = {1096-1099},
       adsurl = {https://ui.adsabs.harvard.edu/abs/1982CSci...51.1096R}
}

@ARTICLE{rea17,
       author = {{Rea}, N. and {Coti Zelati}, F. and {Esposito}, P. and {D'Avanzo}, P. and {de Martino}, D. and {Israel}, G.~L. and {Torres}, D.~F. and {Campana}, S. and {Belloni}, T.~M. and {Papitto}, A. and {Masetti}, N. and {Carrasco}, L. and {Possenti}, A. and {Wieringa}, M. and {Wilhelmi}, E. De O{\~n}a and {Li}, J. and {Bozzo}, E. and {Ferrigno}, C. and {Linares}, M. and {Tauris}, T.~M. and {Hernanz}, M. and {Ribas}, I. and {Monelli}, M. and {Borghese}, A. and {Baglio}, M.~C. and {Casares}, J.},
        title = "{Multiband study of RX J0838-2827 and XMM J083850.4-282759: a new asynchronous magnetic cataclysmic variable and a candidate transitional millisecond pulsar}",
      journal = {\mnras},
         year = 2017,
        month = nov,
       volume = {471},
       number = {3},
        pages = {2902-2916},
          doi = {10.1093/mnras/stx1560},
archivePrefix = {arXiv},
       eprint = {1611.04194},
 primaryClass = {astro-ph.HE},
       adsurl = {https://ui.adsabs.harvard.edu/abs/2017MNRAS.471.2902R}
}

@ARTICLE{ray13,
       author = {{Ray}, P.~S. and {Ransom}, S.~M. and {Cheung}, C.~C. and {Giroletti}, M. and {Cognard}, I. and {Camilo}, F. and {Bhattacharyya}, B. and {Roy}, J. and {Romani}, R.~W. and {Ferrara}, E.~C. and {Guillemot}, L. and {Johnston}, S. and {Keith}, M. and {Kerr}, M. and {Kramer}, M. and {Pletsch}, H.~J. and {Saz Parkinson}, P.~M. and {Wood}, K.~S.},
        title = "{Radio Detection of the Fermi-LAT Blind Search Millisecond Pulsar J1311-3430}",
      journal = {\apjl},
         year = 2013,
        month = jan,
       volume = {763},
       number = {1},
          eid = {L13},
        pages = {L13},
          doi = {10.1088/2041-8205/763/1/L13},
archivePrefix = {arXiv},
       eprint = {1210.6676},
 primaryClass = {astro-ph.HE},
       adsurl = {https://ui.adsabs.harvard.edu/abs/2013ApJ...763L..13R}
}

@ARTICLE{rom12,
       author = {{Romani}, Roger W.},
        title = "{2FGL J1311.7-3429 Joins the Black Widow Club}",
      journal = {\apjl},
         year = 2012,
        month = aug,
       volume = {754},
       number = {2},
          eid = {L25},
        pages = {L25},
          doi = {10.1088/2041-8205/754/2/L25},
archivePrefix = {arXiv},
       eprint = {1207.1736},
 primaryClass = {astro-ph.HE},
       adsurl = {https://ui.adsabs.harvard.edu/abs/2012ApJ...754L..25R}
}

@ARTICLE{rom15,
       author = {{Romani}, Roger W. and {Filippenko}, Alexei V. and {Cenko}, S. Bradley},
        title = "{A Spectroscopic Study of the Extreme Black Widow PSR J1311-3430}",
      journal = {\apj},
         year = 2015,
        month = may,
       volume = {804},
       number = {2},
          eid = {115},
        pages = {115},
          doi = {10.1088/0004-637X/804/2/115},
archivePrefix = {arXiv},
       eprint = {1503.05247},
 primaryClass = {astro-ph.HE},
       adsurl = {https://ui.adsabs.harvard.edu/abs/2015ApJ...804..115R}
}

@ARTICLE{rom16,
       author = {{Romani}, Roger W. and {Sanchez}, Nicolas},
        title = "{Intra-binary Shock Heating of Black Widow Companions}",
      journal = {\apj},
         year = 2016,
        month = sep,
       volume = {828},
       number = {1},
          eid = {7},
        pages = {7},
          doi = {10.3847/0004-637X/828/1/7},
archivePrefix = {arXiv},
       eprint = {1606.03518},
 primaryClass = {astro-ph.HE},
       adsurl = {https://ui.adsabs.harvard.edu/abs/2016ApJ...828....7R}
}

@ARTICLE{san17,
       author = {{Sanchez}, Nicolas and {Romani}, Roger W.},
        title = "{B-ducted Heating of Black Widow Companions}",
      journal = {\apj},
         year = 2017,
        month = aug,
       volume = {845},
       number = {1},
          eid = {42},
        pages = {42},
          doi = {10.3847/1538-4357/aa7a02},
archivePrefix = {arXiv},
       eprint = {1706.05467},
 primaryClass = {astro-ph.HE},
       adsurl = {https://ui.adsabs.harvard.edu/abs/2017ApJ...845...42S}
}

@ARTICLE{sha17,
       author = {{Shahbaz}, T. and {Linares}, M. and {Breton}, R.~P.},
        title = "{Properties of the redback millisecond pulsar binary 3FGL J0212.1+5320}",
      journal = {\mnras},
         year = 2017,
        month = dec,
       volume = {472},
       number = {4},
        pages = {4287-4296},
          doi = {10.1093/mnras/stx2195},
archivePrefix = {arXiv},
       eprint = {1708.07355},
 primaryClass = {astro-ph.HE},
       adsurl = {https://ui.adsabs.harvard.edu/abs/2017MNRAS.472.4287S}
}

@ARTICLE{sir11,
       author = {{Sironi}, Lorenzo and {Spitkovsky}, Anatoly},
        title = "{Acceleration of Particles at the Termination Shock of a Relativistic Striped Wind}",
      journal = {\apj},
         year = 2011,
        month = nov,
       volume = {741},
       number = {1},
          eid = {39},
        pages = {39},
          doi = {10.1088/0004-637X/741/1/39},
archivePrefix = {arXiv},
       eprint = {1107.0977},
 primaryClass = {astro-ph.HE},
       adsurl = {https://ui.adsabs.harvard.edu/abs/2011ApJ...741...39S}
}

@ARTICLE{swi22,
       author = {{Swihart}, Samuel J. and {Strader}, Jay and {Chomiuk}, Laura and {Aydi}, Elias and {Sokolovsky}, Kirill V. and {Ray}, Paul S. and {Kerr}, Matthew},
        title = "{A New Flaring Black Widow Candidate and Demographics of Black Widow Millisecond Pulsars in the Galactic Field}",
      journal = {\apj},
         year = 2022,
        month = dec,
       volume = {941},
       number = {2},
          eid = {199},
        pages = {199},
          doi = {10.3847/1538-4357/aca2ac},
archivePrefix = {arXiv},
       eprint = {2210.16295},
 primaryClass = {astro-ph.HE},
       adsurl = {https://ui.adsabs.harvard.edu/abs/2022ApJ...941..199S}
}

@ARTICLE{str14,
       author = {{Strader}, Jay and {Chomiuk}, Laura and {Sonbas}, Eda and {Sokolovsky}, Kirill and {Sand}, David J. and {Moskvitin}, Alexander S. and {Cheung}, C.~C.},
        title = "{1FGL J0523.5-2529: A New Probable Gamma-Ray Pulsar Binary}",
      journal = {\apjl},
         year = 2014,
        month = jun,
       volume = {788},
       number = {2},
          eid = {L27},
        pages = {L27},
          doi = {10.1088/2041-8205/788/2/L27},
archivePrefix = {arXiv},
       eprint = {1405.5533},
 primaryClass = {astro-ph.HE},
       adsurl = {https://ui.adsabs.harvard.edu/abs/2014ApJ...788L..27S}
}

@ARTICLE{str19,
       author = {{Strader}, Jay and {Swihart}, Samuel and {Chomiuk}, Laura and {Bahramian}, Arash and {Britt}, Chris and {Cheung}, C.~C. and {Dage}, Kristen and {Halpern}, Jules and {Li}, Kwan-Lok and {Mignani}, Roberto P. and {Orosz}, Jerome A. and {Peacock}, Mark and {Salinas}, Ricardo and {Shishkovsky}, Laura and {Tremou}, Evangelia},
        title = "{Optical Spectroscopy and Demographics of Redback Millisecond Pulsar Binaries}",
      journal = {\apj},
         year = 2019,
        month = feb,
       volume = {872},
       number = {1},
          eid = {42},
        pages = {42},
          doi = {10.3847/1538-4357/aafbaa},
archivePrefix = {arXiv},
       eprint = {1812.04626},
 primaryClass = {astro-ph.HE},
       adsurl = {https://ui.adsabs.harvard.edu/abs/2019ApJ...872...42S}
}
\bibliographystyle{aasjournal}

\end{document}